# HPC as a Service: A naïve model

Hamza Ali Imran
EPIC Lab FAST-National University of
Computer & Emerging Sciences
Islamabad, Pakistan
himran.mscs18seecs@seecs.edu.pk

Saad Wazir
EPIC Lab FAST-National
University of
Computer & Emerging Sciences
Islamabad, Pakistan
swazir.mscs18seecs@seecs.edu.pk

Ahmed Jamal Ikram
EPIC Lab FAST-National
University of
Computer & Emerging Sciences
Islamabad, Pakistan
ahmed.jamal.ikram@gmail.com

Ataul Aziz Ikram
EPIC Lab FAST-National University of
Computer & Emerging Sciences
Islamabad, Pakistan
ata.ikram@nu.edu.pk

Hanif Ullah
Research Scholar
Riphah International University
Islamabad, Pakistan
hukhan.dev@gmail.com

Maryam Ehsan
Information Technology
Department
University of Gujrat
Gujrat, Pakistan
maryam.ehsan@uog.edu.pk

*Abstract*—Applications like Big Data, Machine Learning, Deep Learning and even other Engineering and Scientific research requires a lot of computing power; making High-Performance Computing (HPC) an important field. But access to Supercomputers is out of range from the majority. Nowadays Supercomputers are actually clusters of computers usually made-up of commodity hardware. Such clusters are called Beowulf Clusters. The history of which goes back to 1994 when NASA built a Supercomputer by creating a cluster of commodity hardware. In recent times a lot of effort has been done in making HPC Clusters of even single board computers (SBCs). Although the creation of clusters of commodity hardware is possible but is a cumbersome task. Moreover, the maintenance of such systems is also difficult and requires special expertise and time. The concept of cloud is to provide on-demand resources that can be services, platform or even infrastructure and this is done by sharing a big resource pool. Cloud computing has resolved problems like maintenance of hardware and requirement of having expertise in networking etc. An effort is made of bringing concepts from cloud computing to HPC in order to get benefits of cloud. The main target is to create a system which can develop a capability of providing computing power as a service which to further be referred to as Supercomputer as a service. A prototype was made using Raspberry Pi (RPi) 3B and 3B+ Single Board Computers. The reason for using RPi boards was increasing popularity of ARM processors in the field of HPC

## I. Introduction

There is always an increasing demand for powerful compute resources and recent advancements in applications like deep learning, data mining and machine learning etc. have increased this demand a lot. The requirement of increasing demands has been fulfilled by parallel computing for a long time now. Although creation of clusters of commodity hardware is possible but is a cumbersome task. The maintenance of such systems is also difficult and time consuming. Not everyone can afford to buy and maintain such systems. Moreover, we need to have expertise in Linux and networking if we want to make use of such systems. This reduces our focus on programming for actual work. The concept of cloud is to provide on-demand resources that can be services, platform or even infrastructure and this is done by sharing a big resource pool. Cloud computing has resolved the problems like maintenance of hardware and requirement of having expertise in networking etc. This research has made an effort of bringing concepts from cloud computing to HPC in order to get benefits of cloud. The main target is to create a system which can use supercomputer as a service. For this purpose, the criteria include,

- System to literally have any sort of commodity hardware at back end.
- Nodes could be even SBCs like Raspberry Pi, Parallela, Pine64 etc.
- The designed system would hide all complexities of Cluster from end-user, making programmer free from having expertise in Linux and Networking.

- System could be accessible from anywhere in the world so that people living in underdeveloped countries also have access to it.

We started targeting the existing solutions and tried to find what solutions could help us. Our problem could be divided into three major parts. The Cluster formation, which we called backend, the User Interface (UI) formation which we called frontend and finally the interface formation which would join them. For all of these three portions, we used open-source tools. For backend MPICH, MPI4py and Ganglia were used. Moreover, the operating systems used was Raspbian Stretch which is a Debian based Linux distribution. Message Passing Interface (MPI) is becoming a standard for Supercomputing domain. It allows the user to distribute the load over the clusters of any type of computers. It creates processes, which are actually the same copy of the code but each process is assigned a unique ID called its Rank starting from zero. Master process is assigned rank zero. Depending upon the rank number, the programmer decides the job of any particular process. Processes can communicate with each other by passing messages between each other. Applications which can be easily divided into independent chucks are most suitable to run on such clusters. Such applications are called embarrassingly parallel problems. User tells cluster the number of processes to be created at execution time. MPICH is an open-source implementation of message passing interface and is one of the most famous open-source implementations of it. It allows the user to program in C, C++ and FORTRAN. Python is one of the most famous scripting languages and is one of the most commonly used scripting languages for machine learning and data mining applications. Furthermore, it has also become a famous choice in the field of HPC because of its ease in programming. So we used MPI4py which is a Python binding for MPI. Both MPICH and MPI4py were built from source code on Raspberry pi 3B and 3B+. Ganglia is a tool for monitoring distributed HPC clusters. It is the most suitable tool for monitoring a distributed system; similar to the desired tasks that need to be performed. The main task was to create a Web UI (User Interface) which was to be linked to the cluster along with that, it should also be accessible from any device having internet access. Making the end-user able to simply log in to the designed page and upload his or her code and enter the number of processes. The files would be transferred to the cluster, executed and final results would be shown back to the user on the web browser. The open-source tools used for making UI (User Interface) and interfacing it with cluster were PHP, Apache Server, MySQL and Bootstrap. Front end is also designed in a way that it can be deployed on any commodity hardware even on SBCs. The hardware used for creating prototype were Raspberry pi 3B and 3B+ single board computers. The reasons for using them were two. First that we wanted to prove that any type of computer resource could be used for our system. Second reason was, we wanted to use ARM processors. It is estimated that in coming future ARM processors will replace Intel Xeon processors in data centers. The reason is, they are cheap. Hence can be used in abundance and they also produce much less heat then Xeon processors hence reduces the need for Ventilation.

This paper presents a Naive model to provide High-Performance Compute Resource as a service and also presents a working prototype made-up of Cheap less power consuming single board computers which is an extension of our previous work [8].

## II. LITERATURE REVIEW

The integration of SBCs and their applications to the development of cluster and HPC have been a topic of enormous interest since the past few years and it has been observed in the published research.

A study of the architecture as well as the application of the supercomputer was discussed by Atul Gonsai and Bhargavi Goswami [1]. They proposed that a normal system may be changed and upgraded into a supercomputer by altering its configuration employing parallel computing as well as HPC. Abhishek Gupta and Dejan Milojici [2] proposed an evaluation of the performance and cost tradeoffs of HPC applications on a range of platforms varying from cloud to a tailor-made and optimized cluster. The results showed that the cloud was better for platform-specific, non-communication-invasive applications having a low processor count. Pratima Dhuldhule et al [3] presented an idea of using HPC as a platform as a service setup. The architecture presented built an HPC platform by providing a group of nodes that were booted with a desired HPC environment bypassing the virtualization layer by employing technologies like Wake-on-LAN and network booting with the specific functionality of starting, scaling and terminating the computer cluster. In 2016 Abrrachman Mappuji et al. [4] presented a utilization of multiple Raspberry pi 2 SBCs as a cluster to compensate its computing power. The findings were that the increase in every SBC member in a cluster is not necessarily a measure for a significant increase in computational capabilities, and also recommending that 4 nodes are a maximum for an SBC cluster; based for optimum power consumption. Kim and Son [5] proposed a case study on the feasibility of Raspberry-Pi as a big cluster in the smart factory. Appavo et al's publication [6] introduced and analyzed a hybrid supercomputer software infrastructure, to allows direct hardware access to the communication hardware for the necessary components while providing the standard elastic cloud infrastructure for other components. Amazon's EC2 and Microsoft's Azure are examples of cloud computing offerings available today.

Although little is published about the platforms, a trend in the products targeting cloud computing, such as Rackable's MicroSliceTM products [7] is seen, which, with a total of 264 servers per rack, provides a density that is 3 to 6 times the density of a normal rack. Such systems still rely on standard external Ethernet switch infrastructure for connectivity but we predict that the trends in commercial system integration will follow a trajectory similar to that of supercomputers.[9] demonstrate a portable cluster which they called Pi Stack to be used as a platform for computation at edge level. [10] built a micro-scale datacenter using Raspberry Pi and emulate every layer of cloud which can be used to teach cloud computing architecture and infrastructure at a very small scale. [11] favored the implementation of low cost SBC based cluster to perform parallel computation with detail comparison and consideration of cluster construction techniques, Power usage, Scaling and Value for money.

### III. BLOCK AND SYSTEM-LEVEL DIAGRAMS

The Block Diagram of the overall system can be seen in Figure 1. It has majorly two parts. First part is the cluster on the back end; which have one master node and several slave node. Second part is the front end, which actually creates a link between end-user and the cluster. For communication of messages between the nodes, a network is needed, which is further created with a second layer switch and CAT-5 cables.

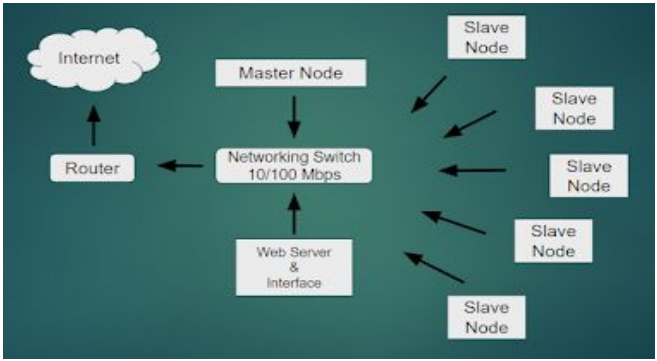

Fig. 1. Block diagram

The main feature of the proposed system is that it is easily extendable i.e more nodes can be added to the cluster. Even the nodes can be Virtual Machine (VM) running on any cloud. Moreover, the system can be globally distributed. The target is to present a proof of concept due to which a limited number of nodes are used. Following is the system diagram (Figure 2) of hardware which is used.

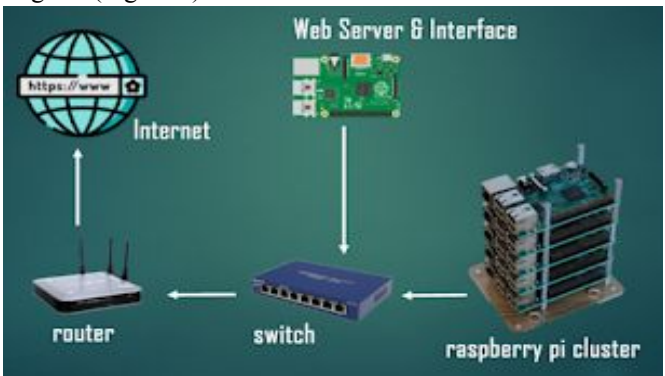

Fig. 2. System Diagram

### IV. FRONT-END: WEB-BASED USER INTERFACE

Front end is designed to provide an easy-to-use graphical user interface from which user has the capability to upload a program file (Parallel code of Python using MPI4py), execute it on a cluster and get the results instantly on a web browser. It is capable of work for any kind of cluster and is also capable of being deployed on any sort of commodity hardware. User Interface (UI) is developed on PHP, MySQL, Apache and Bootstrap.PHP was used as a server-side language, MySQL was handling database and Apache Server was used as a Web-server. The target was to give access to cluster or compute power to any sort of system that can be even a smartphone. For that purpose, Bootstrap is used. It helped in making minimal design and providing multiple device support. UI is totally web-based so it can run from any browser just by entering URL of server on which Web-UI is deployed just like any other website. Front end consisted of the following different modules

- User Management Module
- Configuration Module
- File Module
- Execution Module

#### A. User Management Module

This module control user access and activities. Users get their login and password from the administrator. User enters credentials in the login screen to get access to the dashboard. Once the user accesses the dashboard, users having only administrator privileges can access the configuration panel and have other modules access like File Upload Panel, Files menu and Execution panel. Non- admin user can only access File Upload Panel, Files menu and Execution panel.

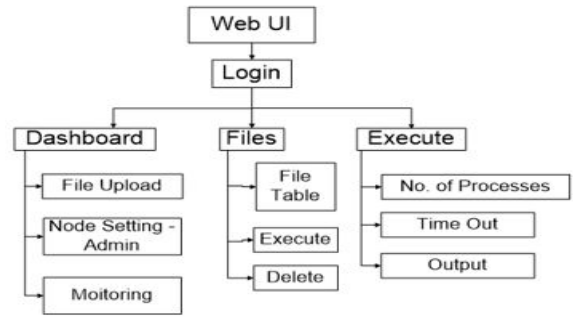

Fig. 3. User Management Module

#### B. Configuration Module

This module is only accessible by users who have administrator privileges. Node Settings page can be opened from the dashboard where the user can add, delete and modify cluster nodes information. The information required to add or modify the cluster node is the IP Address and Username of cluster nodes. After entering all the details of cluster nodes UI is ready to be used. RSA key generated on the front end is to be placed in authorized keys of Master node so that it can SSH in it without a password.

#### C. File Module

This module handles all the files activities which are described stepwise below.
1. User uploads the file from File Upload Panel
2. The file is uploaded from client computer to Web Server
3. File record is maintained in the database
4. File name is changed to preserve its uniqueness
5. A pointer is assigned to the file, which is a number consisting of user id and file no
6. User files can be viewed from Files menu

User can see the details of their uploaded program files in the form of an interactive table and perform delete and execute operations. Procedure is shown in figure 4.

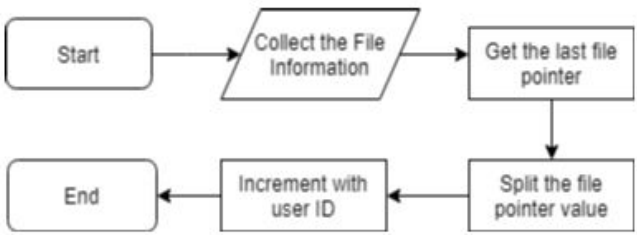

Fig. 4. File Module

### D. Execution Module

Execution module gets the file details from the database and executes the file on the cluster to get the results. A library of PHP named "phpseclib" is used to SCP (Secure Copy Protocol) the program file and SSH (Secure Shell Protocol) into the master node of the cluster. After sending file and getting SSH in to the master, it sends the following commands to the master node
1. SCP the program file to all slave nodes
2. Execute the program by creating N number of processes, where N is the number of processes entered by the user at the time of execution.

Procedure is shown in figure 5.

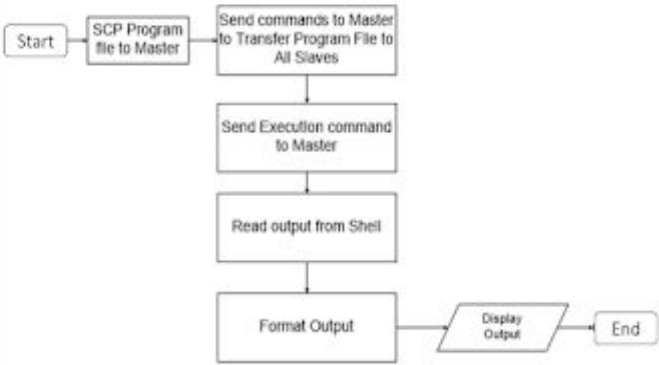

Fig. 5. Execution Module

## V. WORKING

### A. Working of Processes

As previously mentioned, MPI creates processes which are actually the same copy of the code. But each process is assigned a unique process ID called a rank. With the help of rank, we differentiate which process would do which task. Like for process zero (master process), it always has to receive results from all other processes because in the end, it has to print the final result. As the process is the same copy of the code, MPI requires us to store the source code (in case of scripting languages like Python) or compiled file (in case of compiled languages like C++) at the same location and with the same name on all the nodes. The flowchart below (figure 6) explains how any typical process will execute.

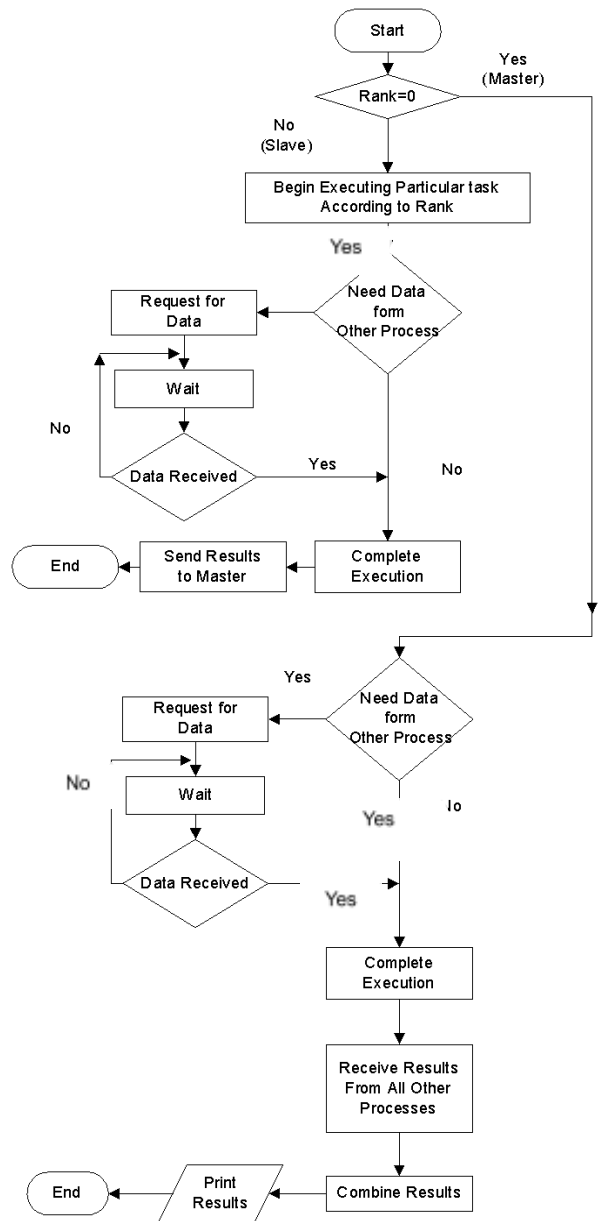

Fig. 6. Working of Process

### B. Working of MPI Cluster

MPI Clusters have master-slave architecture. One node is the master which has control over all the slaves. A cluster can consists to be only a single node and can also have theoretically any number of slaves. In order to give the master the access to all nodes, an RSA key was generated on the master which was then stored as authorized key at all the nodes. Hence making master capable of SSH and send File by SCP protocol without any password. The Basic working of Cluster is explained in the flowchart below (Figure 7).

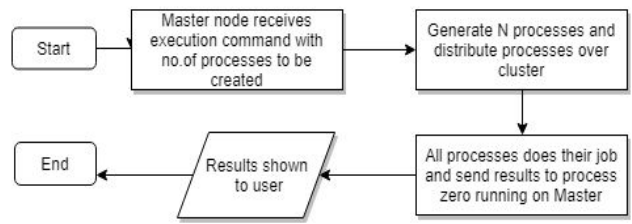

Fig. 7. Working of MPI Cluster

## C. Hierarchy of tools

Following diagram (figure 8) shows the hierarchy of tools used for this project.

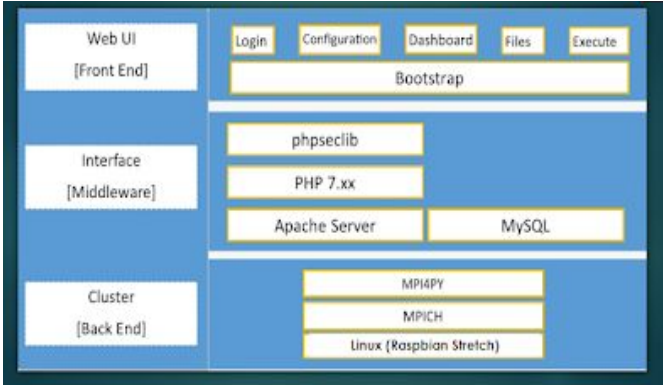

Fig. 8. Hierarchy of tools

## D. Hardware Used

Following Table (Table 1) shows the hardware which was used for this project.

TABLE I. HARDWARE USED

| Sr. No. | Component | Details |
|---|---|---|
| 1 | Web Server & Interface | **Raspberry Pi 3B+** |
| | | **CPU:** 4x 1.4GHz 64-bit quad-core ARM Cortex-A53 CPU |
| | | **RAM:** 1GB LPDDR2 SDRAM |
| | | **Ethernet:** Gigabit Ethernet over USB 2.0 (max 300 Mbps) |
| | | **Memory Card**: Samsung EVO Plus - Class 10 8GB |
| 2 | Master Node | **Raspberry Pi 3B+** |
| | | **CPU:** 4x 1.4GHz 64-bit quad-core ARM Cortex-A53 CPU |
| | | **RAM:** 1GB LPDDR2 SDRAM |
| | | **Ethernet:** Gigabit Ethernet over USB 2.0 (max 300 Mbps) |
| | | **Memory Card**: Samsung EVO Plus - Class 10 8GB |
| 3 | Slave Nodes | **Raspberry Pi 3B** |
| | | **CPU:** 1.2GHz 64-bit quad-core ARM Cortex-A53 CPU |
| | | **RAM:** 1GB LPDDR2 SDRAM |
| | | **Ethernet:** 10/100 Ethernet, |
| | | **Memory Card**: Samsung EVO Plus - Class 10 8GB |
| 4 | Router | **Tenda Wireless N300 Router** |
| | | **Model No. :** 4G630 |
| 5 | Networking Switch | **HP 24 Ported 10/100 Mbps** |

## E. Algorithms Implemented

In order to test our systems working, we have implemented a famous embarrassingly parallel algorithm Monte Carlo's Method for finding the value of Pi. Following is its algorithm.

```
Start_time = Current_time()
Max_tries = 10000000
Total_processes = Get_size()
Each_proc_itr = Max_tries/Total_processes
hits = 0
if myid != 0:# it is a slave
        for i in range (0, Each_proc_itr):
            x = random()
            y = random()
            dist = sqrt(pow(x, 2) + pow(y, 2))
            if dist < = 1.0: # in circle
                hits = hits + 1.0
        Send(hits, destination = 0)
else:# it is master process
        Total_hits = 0
        for i in range (0, Each_proc_itr):
            x = random()
            y = random()
            dist = sqrt(pow(x, 2) + pow(y, 2))
            if dist < = 1.0:
                hits = hits + 1.0
        Total_hits = Total_hits + hits
        for i in range(1,Total_processes):
            recvmsg = Receive( source = i)
            Total_hits = Total_hits + recvmsg
        my_pi = 4 * (Total_hits)/(Max_tries)
        print Total Hits  =   % (Total_hits)
        print Total Tries =   % (Max_tries)
        print Approx pi =   % (my_pi)
        print Time (ms) = % (Current_time() - Start_time)
```

## F. Experimental Results

The above algorithm was executed by creating a different number of processes. Time taken was measured in milliseconds. A tabular (Table 2) as well as graphical representation (Figure 9) of the data is shown.

TABLE II: TIME TAKEN VERSUS NUMBER OF PROCESSES

| No of Processes | Time (ms) |
|---|---|
| 1 | 65758.236 |
| 2 | 34246.604 |
| 4 | 33053.88 |
| 6 | 22656.33 |
| 8 | 17804.702 |
| 10 | 14243.421 |
| 12 | 12387.994 |
| 14 | 10648.292 |
| 16 | 9491.039 |
| 18 | 8633.861 |
| 20 | 7873.781 |
| 22 | 7245.755 |
| 24 | 6868.603 |
| 26 | 6397.734 |
| 28 | 5965.643 |
| 30 | 5721.591 |

| 32 | 5467.573 |

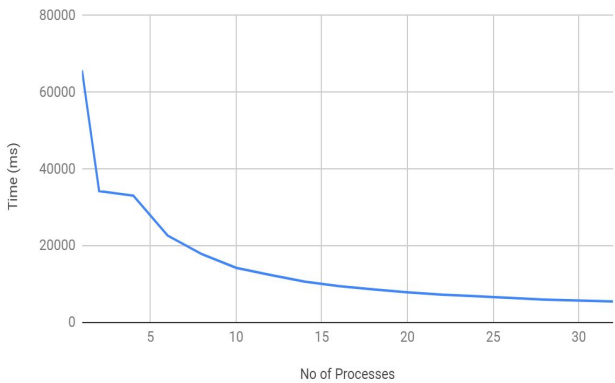

Fig. 9. Time taken Versus Number of Processes

### G. Screenshots of UI

Following (Figure 10, 11 & 12) are some screenshots of the designed and developed UI.

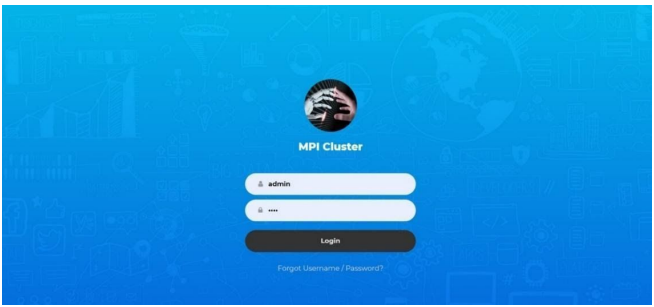

Fig. 10. Login Page

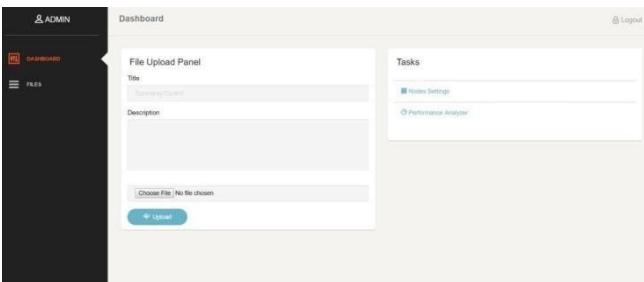

Fig. 11. File Upload Page

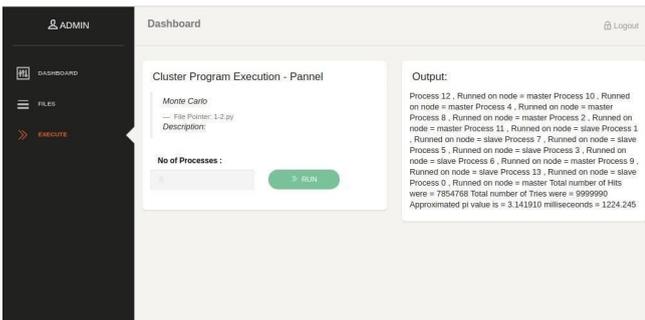

Fig. 12. Execution and Output Page

## VI. CONCLUSIONS AND RESULTS

The access to High-Performance Compute Resource is out of range from the majority. We have made an effort to make a system using open-source tools to create a link between High-Performance Compute Resource and people who need it but could not access it or maintained it before. Everyone has access to the internet since Smartphone became cheap and easily available. Using our system; HPC can be made accessible from Smartphone even.

## VII. FUTURE WORK

In future text editor can also be added in the Web UI. Moreover, currently, our system is only having support of MPI4py (Python). Other Languages like Perl, Octave for which MPI bindings are available; support can also be added. The support of uploading a code of compiled languages like C or C++ and compiling code on a cluster can also be added.